\documentclass[twocolumn,english,aps,prl,showpacs]{revtex4}
\usepackage{graphicx}
\usepackage{bm}
\begin{document}

\title{Breakdown of the power-law decay prediction of the heat
current correlation in one-dimensional momentum conserving lattices}
\author{Shunda Chen}
\author{Yong Zhang}
\author{Jiao Wang}
\author{Hong Zhao}
\email{zhaoh@xmu.edu.cn}
\affiliation{Department of Physics and Institute of Theoretical Physics
and Astrophysics, Xiamen University, Xiamen 361005, Fujian, China}
\begin{abstract}
We show that the asymmetric inter-particle interactions can induce
rapid decay of the heat current correlation in one-dimensional momentum
conserving lattices. When the asymmetry degree is appropriate, even
exponential decay may arise. This fact suggests that the power-law
decay predicted by the hydrodynamics may not be applied to the
lattices with asymmetric inter-particle interactions, and as a result,
the Green-Kubo formula may instead lead to a convergent heat conductivity
in the thermodynamic limit. The mechanism of the rapid decay is traced
back to the fact that the heat current has to drive a mass current
additionally in the presence of the asymmetric inter-particle interactions.
\end{abstract}

\pacs{05.60.Cd, 44.10.+i, 66.70.-f, 63.20.-e}
\maketitle

Onsager's classic work \cite{Onsager} has shown that in the linear
response region, the current $J_{i}$ of a physical quantity can be
expanded in terms of the so-called thermodynamic forces,
i.e., $J_{i}=\sum_{j}\kappa_{ij}F_{j}$. The Onsager coefficient
$\kappa_{ij}$, which describes the coupling between the forces $F_{i}$
and $F_{j}$, can be calculated in terms of the correlation function
of the spontatous current fluctuations as \cite{KuboSP},
\begin{equation}
\kappa_{ij}=\lim\limits _{\tau\rightarrow\infty}\lim\limits _{L\rightarrow\infty}\frac{1}{2}\int_{0}^{\tau}C_{ij}(t)dt,
\end{equation}
where $C_{ij}(t)\equiv \langle (J_{i}(t)-\langle J_{i}\rangle)
(J_{j}(0)-\langle J_{j}\rangle)\rangle$ and $L$ is the linear dimension
of the system along which the current flows.
The celebrated Onsager reciprocal relation formulates the relation
between the coupling coefficients, which gives $\kappa_{ij}=\kappa_{ji}$.
However, we recall that under what condition the coupling coefficients
are non-vanishing, and what a role the couplings may play in characterizing
the transport process, have not been clarified.

Equation (1) is the well-known Green-Kubo formula. It provides a way
for calculating the transport coefficient by considering the current
fluctuations in equilibrium state. The traditional hydrodynamic approach
assumed that the current correlation decays rapidly, i.e., exponentially
fast, so to ensure the convergence of the integral in the Green-kubo
formula \cite{TheorySP,Nara}. However, after Alder numerically evidenced
the `long time tail' of the correlation function of the energy current
in 1970 \cite{Alder} in a gas model, a lot of theoretical analysis as
well as numerical simulations have shown that the autocorrelation
functions of currents in one-dimensional (1D) fluids may generally decay
in a power-law manner instead \cite{Lebowitz,Lepri,Dharrev}. The power-law
decay is explained within the framework of hydrodynamics, where the slow
diffusion of long wave hydrodynamic modes or the ring-collision of
particles are ascribed to be the underlying mechanisms.

In recent decades, low-dimensional materials such as nanowires and
graphene flakes have come into focus in many disciplines. The studies
of them are undergoing rapid progress for both fundamental research
interests and various intriguing applications. The heat transport properties
of low-dimensional lattice models have been a particularly interesting
issue. An important progress is that the hydrodynamic analysis has been
extended to the study of the transport behavior in lattices. Based on
intensive theoretical analysis \cite{Pros, Mai, Nara, Beijeren12, JSW,
Del, DelStat} and numerical studies (see for example \cite{Lepri,
Dharrev} and references there in), for 1D momentum conserving fluids
and lattices, at present it is generally believed that the current
correlation should decay in the power-law manner. An important consequence
of the power-law decay is that the integral in the Green-Kubo formula will
diverge. For a finite system size, to truncate the integration at a reasonable
time may lead to a finite transport coefficient, but, however, it may
diverge in the thermodynamical limit in a way of $\kappa_{ij}\sim L^{\nu}$
(with $\nu>0$).

Meanwhile, several counterexamples to the divergent heat conductivity have
also been reported, including the rotator model \cite{Giar}, a 1D lattice
in an effective magnetic field \cite{Giar2005}, the variant ding-a-ling
model \cite{DaL}, and lattice models with asymmetric inter-particle interactions \cite{Zhong}. We notice that in the counterexamples studied in Ref. \cite{DaL}
and \cite{Zhong}, asymmetric inter-particle interactions are quite relevant.
Hence a natural question is what effect the asymmetric interactions may have
on the `long time tail' decay and result in a finite, convergent heat
conductivity. In fact, lattice models with asymmetric interactions have been
studied in the literature, both analytically \cite{Pros, DelStat, Beijeren12}
and numerically \cite{Pros,FPUAB}. In particular, the hydrodynamic analysis
based on the Burgers equation suggests that $\alpha=1/2$ for systems with
asymmetric interactions and $\alpha=1/3$ for those with only symmetric
interactions \cite{Beijeren12}, which agrees with the result based on
the Zwangzig-Mori equation and the self-consistent mode coupling theory
\cite{DelStat}.

\begin{figure}
\includegraphics[width=1.05\columnwidth]{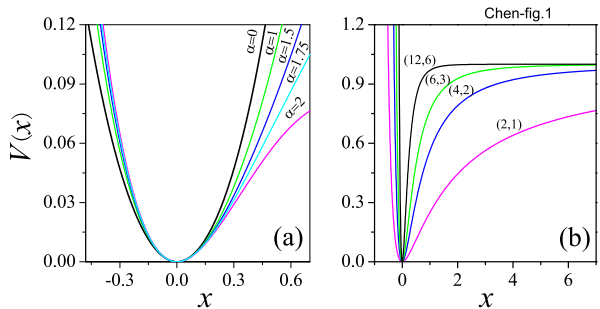}
\vskip-0.4cm
\caption{Plots of the inter-particle potentials investigated: The potential
of (a) the FPU-$\alpha$-$\beta$ lattice model and (b) the L-J lattice model.}
\end{figure}

In this paper, by employing two paradigmatic lattice models, the Fermi-Pasta-Ulam-$\alpha$-$\beta$ (FPU-$\alpha$-$\beta$) model and the
Lennard-Jones (L-J) model, we show that the power-law decay of the current
correlation may vanish in the presence of the asymmetric interactions. Instead,
rapid decay fast than the power law is observed in both models.
It is well known that systems with symmetric and asymmetric interactions have
remarkable different properties. For example, the thermal expansion effect is
exclusive in the latter. In the following, we shall conjecture that the thermal
expansion effect is also responsible for the rapid decay of current correlation functions. For this aim we have to confront the previous numerical results
in the literature and the contradictory hydrodynamic predictions. On one hand,
we shall provide careful, large-scale simulation results, and show the importance
to keep the lattice features of the models in the simulations and the sensitive
dependence of the results on the asymmetry degree. As to the hydrodynamic
predictions, we think it is still open. As the hydrodynamic analysis is based
on the linearized hydrodynamic equations, one possibility is that the role of
the asymmetric interactions may not have been sufficiently considered if their
main effects are contained in the high order terms. We shall show that the
energy current can excite the mass current in systems with asymmetric
interactions, while there lacks such a coupling between the energy and
the mass current in systems of symmetric interactions. We shall also illustrate
that, different from the fluid models, the lattice structure can additionally
scatter the currents. These facts are crucial for the rapid decay of current
correlation but have not been involved explicitly in the previous hydrodynamic
analysis.

Our models are defined by the Hamiltonian
\begin{equation}
H=\sum_{i}\frac{p_{i}^{2}}{2}+V(x_{i}-x_{i-1}-1),
\end{equation}
where $p_{i}$ and $x_{i}$ are the momentum and position of the
$i$th particle, respectively, and $V$ the potential between neighboring
particles. The component particles are assumed to be identical and
have unit mass, and the lattice constant is set to be unity so that
the system length $L$ equals the particle number $N$. The inter-particle
interactions in the FPU-$\alpha$-$\beta$ model is
\begin{equation}
V(x)=\frac{1}{2}x^{2}-\frac{\alpha}{3}x^{3}+\frac{1}{4}x^{4},
\end{equation}
where the parameter $\alpha$ controls the degree of asymmetry as
illustrated in Fig. 1(a). For $\alpha=0$ the system reduces to the
Fermi-Pasta-Ulam-$\beta$ (FPU-$\beta$) model with symmetric potential only.
To well reveal the effects of the asymmetry, in our simulations the
average energy per particle, denoted by $\varepsilon$, is fixed to
be $\varepsilon=0.1$ such that the averaged potential energy per
particle is about $0.05$.

The potential of the L-J model is
\begin{equation}
V(x)=[(\frac{1}{x+1})^{m}-2(\frac{1}{x+1})^{n}+1].
\end{equation}
This potential may well approximate the inter-particle interactions in
many real materials, and hence has important practical implications. In
this potential the degree of asymmetry is controlled by the parameter
set $(m,n)$. (See Fig. 1(b) for several examples.) It should be noted
that if the average potential per particle is larger than one then
the potential models fluids instead. So in our simulations the average
energy per particle is fixed to be $\varepsilon=0.5$ to make sure that
our model is a lattice. As shown in Fig. 1(b), the degree of asymmetry
increases as $(m,n)$ changes from $(12,6)$ to $(6,3)$, and to $(2,1)$.

The energy current \cite{Dhar} is defined as
$J_{q}\equiv\sum_{i}\dot{x}_{i}\frac{\partial V}{\partial x_{i}}.$
For a lattice the energy current is equal to the  heat current because
there is no residual global velocity \cite{Lepri}. To numerically
measure the current in the equilibrium state, the system is first
evolved from an appropriately assigned random initial condition for
a long enough time ($>10^{8}$) to ensure that it has relaxed to the
equilibrium state; then the current at ensuing times is measured. The
periodic boundary condition is applied in the calculations.

\begin{figure}
\includegraphics[width=1.05\columnwidth]{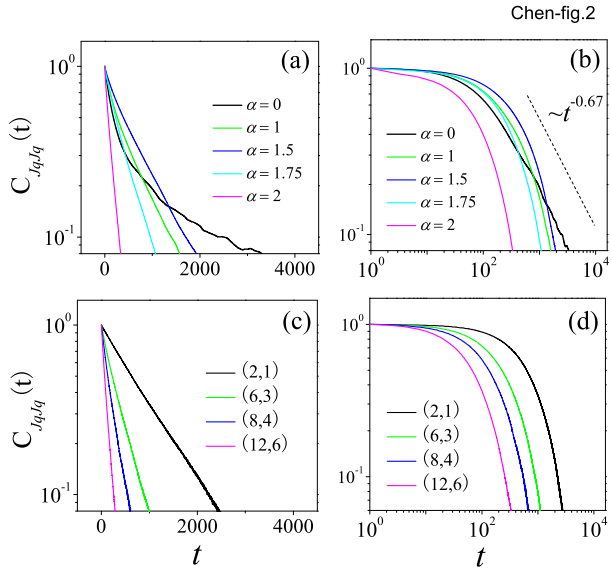}
\vskip-0.4cm
\caption{The autocorrelation function of the heat flux. (a) and (b)
are respectively the semi-log plot and log-log plot of
FPU-$\alpha$-$\beta$ model; (c) and (d) are the same but for the L-J model.}
\end{figure}

Figure 2 shows the autocorrelation of the energy current
$C_{JqJq}(t)\equiv\frac{\langle J_{q}(t)J_{q}(0)\rangle}{\langle J_{q}(0)J_{q}(0)\rangle}$ for the two models. The results are presented
in semi-log scale in panel (a) and (c) and in log-log scale in panel
(b) and (d). In generating Fig. 2, the system size is fixed to be $N=2048$
which is sufficiently large because we have checked that the results does
not change if the system size is increased further. It can be seen that
in the FPU-$\alpha$-$\beta$ model which involves symmetric potential,
the correlation function decays in a power law $C_{J_q J_q}(t)\sim t^{-\gamma}$
with $\gamma\sim 0.67$, which agrees well with previous studies \cite{Dhar}.
Therefore, it provides an example that with the symmetric interactions, the long-time-tail prediction applies. However, for the FPU-$\alpha$-$\beta$
model with $\alpha>1$, we can see that the decay become faster than the
power-law manner which can be roughly regarded to be exponentially. For
the L-J model, the decay is much close to a perfect exponential manner
for different sets of control parameters $(m,n)$. With the increase of
the asymmetry, the decay become faster and faster.

The thermal expansion effect induced by the asymmetric interactions
could be a key for understanding the mechanism \cite{Zhong}. Thermal
expansion implies the coupling between the energy and mass distribution.
Because of the coupling, redistribution of energy can induce redistribution
of mass. In the following, we demonstrate that in equilibrium state,
asymmetry interactions can do result in the coupling between the
fluctuations of the energy current and the mass current.  As the global
mass current in equilibrium state is zero due to the fact that the total
momentum of the system is zero, the cross correlation between the global
energy and mass currents is zero as well. However, the local mass current
of a part of the system, i.e., $J_{m}\equiv Mv$, may still fluctuate.
Here $M$ is the mass of the whole part of the system and $v$ represents
its center-of-mass velocity. We thus investigate the coupling between
the local energy current and the local mass current. To be concrete, in
the simulations we consider a quarter of the whole system as our local
subsystem.

\begin{figure}
\includegraphics[width=1.05\columnwidth]{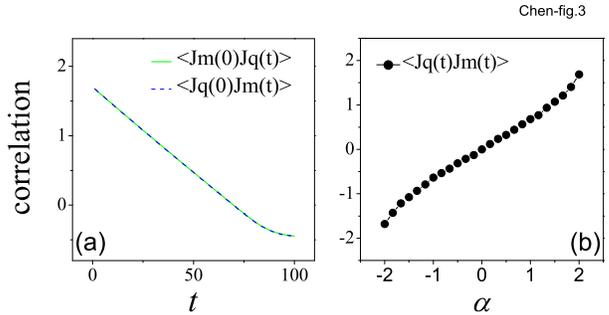}
\vskip-0.4cm
\caption{(a) The cross correlation functions between the local heat flux
$Jq$ and the local mass flux $J_{m}$ for the FPU-$\alpha$-$\beta$ model
with $\alpha=2$. (b) The dependence of the cross correlation function
$\left\langle J_{q}(0)J_{m}(0)\right\rangle $ on the asymmetry parameter
$\alpha$.}
\end{figure}

Figure 3(a) shows the cross correlation of the local energy current
$J_{q}$ and the mass current $J_{m}$ as function of time for the
FPU-$\alpha$-$\beta$ model with $\alpha=2$. It can be seen that
$\left\langle J_{q}(0)J_{m}(t)\right\rangle$ and $\left\langle J_{m}(t)J_{q}(0)\right\rangle$ perfectly agree with each other,
suggesting that the Onsager reciprocal relation, $\kappa_{qm}=\kappa_{mq}$,
holds even for the subsystem.  In Fig. 3(b), the cross correlation
at the same time, i.e., $\left\langle J_{q}(t)J_{m}(t)\right\rangle$,
is shown as a function of the asymmetry parameter $\alpha$. It can
be seen that the cross correlation decreases with the decrease of
the asymmetry. In particular, in the symmetric limit with $\alpha=0$,
the correlation vanishes, which reveals clearly that it is the asymmetry
that sustains the coupling between the currents.

The fact that a symmetric potential leads to a vanishing cross current
correlation can be shown analytically. In the continuum limit, the
local energy current can be written as $j=\langle\dot{x}\frac{\partial V}
{\partial x}\rangle$. In addition, the mass density is a scalar variable
and satisfies $\rho(x)=\rho(-x)$ because the potential is symmetric and
hence we have $V(x)=V(-x)$. Considering all of these, it is
straightforward to derive that
\[
\langle\int_{-l}^{l}\dot{x}\frac{\partial V(x)}{\partial x}\dot{x}\rho(x)dx\rangle=-\langle\int_{-l}^{l}\dot{x}\frac{\partial V(x)}{\partial(x)}\dot{x}\rho(x)dx\rangle,
\]
 which implies a zero cross correlation immediately. Here the integration
is taken over the subsystem on which the local mass current is defined.
Here $(-l,l)$ represents the lattice segment of the subsystem.

Therefore, in a system with the asymmetric interactions, once a fluctuation
of energy current forms spontaneously, a mass density fluctuation will
be excited. The resultant mass current is driven by the energy current
due to their nonzero cross correlation, hence the mass current should
serve as a resistant factor to the energy current.

This mechanism alone, however, can not explain the rapid decay of
the current correlation. The conversional hydrodynamic theory has
predict that low-dimensional fluids should show slow-decay of
current correlation. For the 1D diatomic gas previous simulations
\cite{Grassberger} have shown the power-law decay of $C_{qq}(t)
\sim t^{-0.67}$. We have reperformed the simulation and obtain the
same result. In this model, particles interact via the hard-core
collisions, which is the limit case of the asymmetry potential.
This fact indicates that the lattice feature should also be crucial
to the fast decay.

The spatiotemporal cross correlation of fluctuations of the local
energy current and the local mass current may shed light on the role
of the lattice feature. For this purpose we divide the whole system
into $N$ equal bins. Then the global current $J$ can be expressed
as the sum of the local ones, i.e., $J=\sum_k {j_{k}}$, where $j_{k}$
represents the local current of the $k$th bin. We use
\begin{equation}
C(x,t)=\langle j_{k}^{q}(0)j_{l}^{m}(t)\rangle
\end{equation}
to measure the spatiotemporal cross correlation between the local
energy current $j_{k}^{q}$ in the $k$th bin at $t=0$ and the local
mass current $j_{l}^{m}$ in the $l$th bin at time $t$, where
$x\equiv(l-k)$. Figure 4 shows the results for the
FPU-$\alpha$-$\beta$ model with $\alpha=2$ and $\varepsilon=0.1$
and the diatomic gas model with mass ratio 1/3 and and $\varepsilon=1$.
The difference is remarkable: In the lattice model the positive
correlation peaks are followed by regions with negative correlation,
while in the gas model there are only positive correlation peaks. These
observations indicate that in the lattice model, as the mass current
flows in one direction, a companying mass current in the reverse direction
will be excited as well. Because of the coupling between the energy
and mass current, it also implies that an energy current in the reverse
direction is induced as well. In other words, the currents are reflected
as they flow forwards, and a decay mechanism is thus resulted in.
However, such a mechanism is absent in the gas model.

\begin{figure}
\includegraphics[width=1.05\columnwidth]{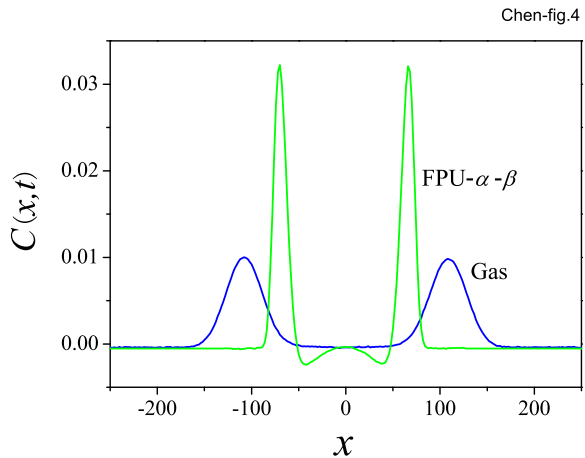}
\vskip-0.4cm
\caption{The spatiotemporal cross correlation of fluctuations of the local
energy current and mass current for (a) the FPU-$\alpha$-$\beta$
model (Green) and (b) the gas model (Blue).}
\end{figure}

In summary, we have shown that the asymmetric inter-particle
interactions could result in nonvanishing coupling between the local
energy current and the local mass current, and the coupling coefficients
follow the Onsager reciprocal relation. In the case of the symmetric
interactions, the coupling vanishes. The asymmetric interactions can
induce rapid decay of the current correlation in 1D momentum conserving
lattices, and results in finite thermal conductivities in the thermodynamic
limit. As the asymmetric interactions, signalized by the thermal expansion
effect \cite{note}, are common in real materials, this finding should
have universal implications.

Our simulations indicate that even perfect exponential decay can be
observed with appropriate degree of asymmetry. However, based on numerical
studies with finite-size effects, at present it is difficult to conclude
with certainty that the asymmetric interactions can $generally$ lead to
the faster than the power-law decay in the current correlation. The degree
of asymmetry is not only determined by the control parameters but also
depends on the temperature of the system. For example, at the same
control parameter of $\alpha=1$ we have observed fast decay of the
current correlation in the FPU-$\alpha$-$\beta$ model, while in previous
studies \cite{FPUAB} a power-law like decay was observed. The difference
roots in the temperature: in \cite{FPUAB} a much higher temperature was
investigated, hence the potential is in effect dominated by the quartic
term and shows a more symmetric structure. To clarify in detail the
dependence of the decay behavior of current correlation on the degree of
asymmetry should be a task for next studies.

Another more challenging task for future studies is to answer why
the direct simulation results as presented here disagree with certain
hydrodynamic predictions. One clue we provide in this paper is that
the coupling effect between the heat and the mass current. With the
asymmetric interactions, local heat currents may excite local mass
currents, hence the latter may be an resistant factor to the former.
In the case of symmetric interactions, the coupling effect vanishes.
On the other hand, the currents can also be scattered by the lattice
structure in the presence of asymmetric interactions. This mechanism
is absent in the fluid model we studied. This observation may
explain why in the one-dimensional diatomic gas a power-law decay
can be observed though the inter-particle interactions are also asymmetric.


This work is supported by the NNSF (Grants No. 10805036, No. 11275159,
and No. 10925525 ) and SRFDP (Grant No. 20100121110021) of China.


\begin{thebibliography}{10}
\bibitem{note}  We have also studied the models that are negatively
thermal expansive, such as in the case of the FPU-$\alpha$-$\beta$
model with $\alpha=-2$, and obtained qualitatively the same results.
\end{thebibliography}
\end{document}